# SPACE CLIMATE MANIFESTATION IN EARTH PRICES – FROM MEDIEVAL ENGLAND UP TO MODERN USA

L.A. PUSTILNIK[1,2], G. YOM DIN[3]
*[1]Israel Cosmic Ray and Space Weather Center, Tel Aviv University and Israel Space Agency, P.O.Box 2217, Katzrin, 12900, Israel; levpust@post.tau.ac.il*
*[2]Sea of Galilee Astrophysical Observatory, Jordan Valley College, 15132, Israel*
*[3]Golan Research Institute, Katzrin, 12900, Israel; rres102@research.haifa.ac.il*

**Abstract.** In this study we continue to search for possible manifestations of space weather influence on prices of agricultural products and consumables. We note that the connection between solar activity and prices is based on the causal chain that includes several nonlinear transition elements. These non-linear elements are characterized by threshold sensitivity to external parameters and lead to very inhomogeneous local sensitivity of the price to space weather conditions. It is noted that "soft type" models are the most adequate for description of this class of connections. Two main observational effects suitable for testing causal connections of this type of sensitivity are considered: burst-like price reactions on changes in solar activity and price asymmetry for selected phases of the sunspot cycle.
The connection, discovered earlier for wheat prices of Medieval England, is examined in this work on the basis of another 700-year data set of consumable prices in England. Using the same technique as in the previous part of our work (Pistilnik and Yom Din 2004) we show that statistical parameters of the interval distributions for price bursts of consumables basket and for sunspot minimum states are similar one to another, like it was reported earlier for wheat price bursts. Possible sources of these consistencies between three different multiyear samples are discussed.
For search of possible manifestations of the 'space weather - wheat market' connection in modern time, we analyze dynamics of wheat prices in the USA in the twentieth century. We show that the wheat prices revealed a maximum/minimum price asymmetry consistent with the phases of the sunspot cycle. We discuss possible explanations of this observed asymmetry, unexpected under conditions of globalization of the modern wheat market.

## 1. Introduction

A history of studying possible connections between space weather and storms in earth markets numbers more than 300 years (Jonathan Swift, 1726; William Hershel, 1801; William Jevons, 1878). In the last years, this problem received a new impulse caused by the discovery of causal connections between the cosmic ray flux penetrated in the Earth atmosphere, and cloudiness (Svensmark and Friis-Christensen, 1997).

In our previous research (Pustilnik and Yom Din, 2004), we reconsidered possible causal connections between solar activity and wheat prices. It was shown that a complex causal chain can have taken place. This chain includes a number of elements, while its basis is the influence of the solar activity on the weather state caused by modulation of galactic cosmic rays propagated into the Solar System to the Earth and their penetration into the earth's atmosphere. As it follows from the study of Svensmark and Friis-Christensen (1997), ions and radicals in the air formed by cosmic rays can be considered as one of essential factors of vapor condensation and cloud formation; their modulation can lead to corresponding variations of the earth weather[1].

1 We wish to point here that the statement about the connection 'cosmic radiation/cloud cover' is a controversial subject (see discussion in studies of Laut (2003), Palle and Butler (2002), Tsiropoula (2003).

From the other side, these weather abnormalities can lead to drops of agricultural production in regions of high risk agriculture, with corresponding market reactions in the form of price bursts. As a result, a causal chain between solar activity and prices of agricultural products can be presented as a sequence of a number of elements (Fig. 1).

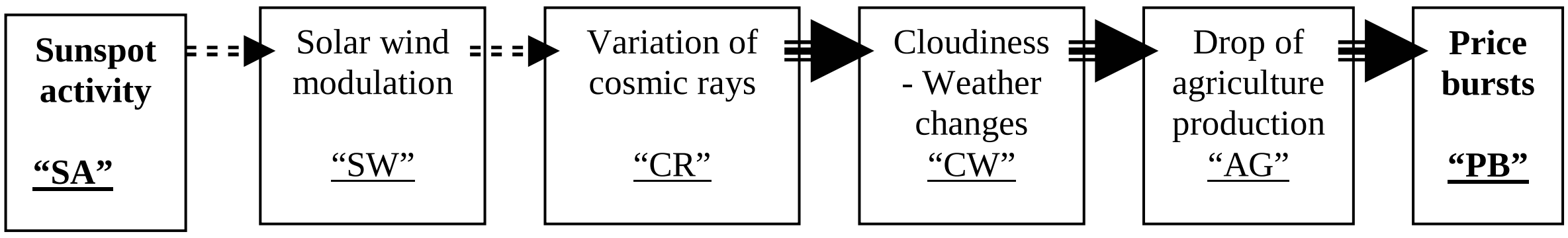

*Figure 1.* A possible causal connection between space weather and price bursts.

A main feature of this chain is a nonlinear type of sensitivity of the last elements of this scheme (marked by large arrows with solid lines). It may lead to step-like transition processes when a small variation of input parameters causes a catastrophic-like transition of the whole system. For example, cloudiness formation caused by the vapor condensation in presence of ions and radicals generated by cosmic rays is very sensitive to vapor concentration, temperature and pressure[2]. This explains a quite inhomogeneous geographical distribution of the sensitivity of "CW" to "CR" (Fig. 1), with a few local spots of very high correlation and extended regions of low correlation (see Fig.8 in Fastrup et al. 2000). Another element with nonlinear sensitivity is a "CW"-"AG" transition that reflects the reaction of agricultural production on weather abnormalities. This process takes place only in regions of high risk agriculture sensitive to weather conditions. Additional parameters here are an agricultural crop and its dynamical range of sensitivity to weather (oat, for example, is more resistant to weather disturbances than wheat and less sensitive to the "CW"-"AG" transition). The market state, in turn, is very sensitive to the last transition element shown in Fig. 1 (price burst reaction on deficit or excess of agricultural production). We wish to note here that existence of reserves and access to external markets with low transfer costs (global market) will suppress the market sensitivity to disturbances of local supply.

As a result, the multi-element chain of the causal connection (Fig. 1) can not be described by "hard type" models with univocal relations like $Y = kX + Noise$, or more generally, $Y^{(n)} = \sum k_i X^i + Noise$, where $X_i$ - input variables (space weather parameters, conditions in the Earth atmosphere, market characteristics,…), $Y$ - the output reaction (market prices, social outcomes, famines), $k_i$ - the coefficients of connections, $^{(n)}$ - the order of derivatives.

On the opposite, this multi-element chain requires "soft type" models for its description when the coefficients of connections $k_i$ depend on the input variables $X_i$ and the output reaction *Y*:

$$Y^{(n)} = \sum k_i(X,Y) * X^i + Noise.$$

---

[2] This explains rigorous requirements to vapor condensation, pressure and temperature, necessary to cosmic ray tracks observations in Wilson camera - analog of cloud formation. Deficit of vapor leads to absence of condensation at all, but if vapor concentration is too large, condensation will have place all the time, independently of external factors and additional input caused by cosmic rays.

This situation is typical for "catastrophy theory" (Arnold 1992) and requires including into consideration hidden parameters of the system. The system's behavior is very sensitive to its location in the multi-dimensional space of $X_i$.

Regarding the problem of space weather influence on earth prices, the soft type of the relationship leads to high sensitivity of this connection to the following important parameters: distribution of vapor in the earth surface determined by global climate and atmospheric circulation; resistance of the agricultural production to weather conditions (crops and their varieties, agro technique and genetics); active participation of the local market in the globalization process (due to cheap shipping costs and low customs). Since all these parameters are very inhomogeneous in space and vary in time on the scale of hundreds of years we can expect that the sensitivity of the market to space weather will be unstable. This sensitivity can take place from time to time in specific regions, when and where all these parameters (density of vapor in atmosphere, state of high risk agriculture, market isolation and restricted external supply) will have occurred simultaneously in one region.

The market behavior expected according to the presented scheme (Fig. 1) has to demonstrate the following two types of reactions on the space weather state:
1. The burst-like price reaction on the crucial combination of the above-considered important parameters. These price bursts are most probable in specific phases of solar activity (minimal or maximal sunspot number) that lead to the most unfavorable states of weather for concrete agricultural crops under concrete local market conditions. Possible types of market reactions were discussed in details in (Pustilnik and Yom Din, 2004) and presented in Fig. 3, 4 in that work.
2. Min/Max price asymmetry - systematical differences between prices in minimum and maximum states of solar activity, caused by the opposite sign of space weather influence on the market in these opposite states of solar activity.

For analysis of concrete situations we have to take into account that global atmospheric circulation that transfers clouds from their birth region to thousands kilometers away (for example, from North Atlantic to East Siberia) may lead to a time lag in weather sensitivity to cosmic ray/sunspot activity, in spite of the vapor state being far from critical in these distanced regions. Another factor of possible increase in system sensitivity to space weather is compactness of agricultural production zones. Clearly, regional sensitivity of crops to weather conditions is much stronger for those of them that are localized in small and compact regions (hundreds kilometers) than for those dispersed on thousands of kilometers (where average weather variations are much smaller).

On the basis of this description we can conclude that standard methods of statistical inference (regression/correlation, Fourier analysis) may be ineffective for the search of the "space weather-price level" connection. Identification of space weather manifestations through Earth markets requires application of another approach based on the event statistics. As it was shown in the previous part of our work (Pustilnik and Yom Din, 2004), adequate methods for this purpose can include (a) statistical study of time intervals between price bursts and (b) search of price asymmetry. Application of this approach to the isolated wheat market of Medieval England has shown the existence of space weather influence on prices both for price burst statistics and for price asymmetry. At the same time, for more reliability we need to test this fact on other independent samples of prices for the same historical period. Another side of the problem is possible manifestations of the "space weather - market state" connection for modern conditions, when market globalization and increased agriculture resistance to unfavorable weather conditions obviously can diminish the weather influence on prices.

## 2. Sources of Data

To test our assumptions about the influence of solar activity on prices we used the following two additional databases of prices:
(a) The first is the Composite Unit of Consumables (CUC) in England for seven centuries, 1264-1954 (Brown and Hopkins, 1956). In this database, the 'farinaceous' item includes wheat, rye, barley, peas, and, in the twentieth century – wheat and potatoes. The item 'farinaceous' constitutes 20% of the CUC and wheat constitutes 37-49% of farinaceous. The CUC is expressed as an index (CUC for the years 1451-1475 = 100). Main sources of data for wheat prices in the study of Brown and Hopkins were Rogers (1887) and Beveridge (1939), and from the beginning of the nineteenth century, wholesale prices on organized markets. Since the contribution of wheat to the CUC is less than 10%, data from the CUC and data on wheat prices are independent of one another for our purposes.
(b) As the second data set for testing efficacy of the proposed causal chain under conditions of the modern wheat market we used the USDA (2004) database that contains average yearly prices in US$ per bushel (Fig.4) received by farmers in the USA for wheat (durum, spring, winter, other kinds, total).

## 3. Results and Discussion

### 3.1. EFFECTS OF SUNSPOT ACTIVITY ON THE CUC

The dynamics of CUC for about 700 years is shown in the upper part in Fig. 2. This period includes several major global-affecting changes in socio-economic conditions (Columbus' discovery of America, continental wars, World Wars I and II). We chose to analyze only part of the available CUC prices, namely 1260-1720 (Fig 2, bottom part), as that was the basis for the first part of our research (Pustilnik and Yom Din, 2004).

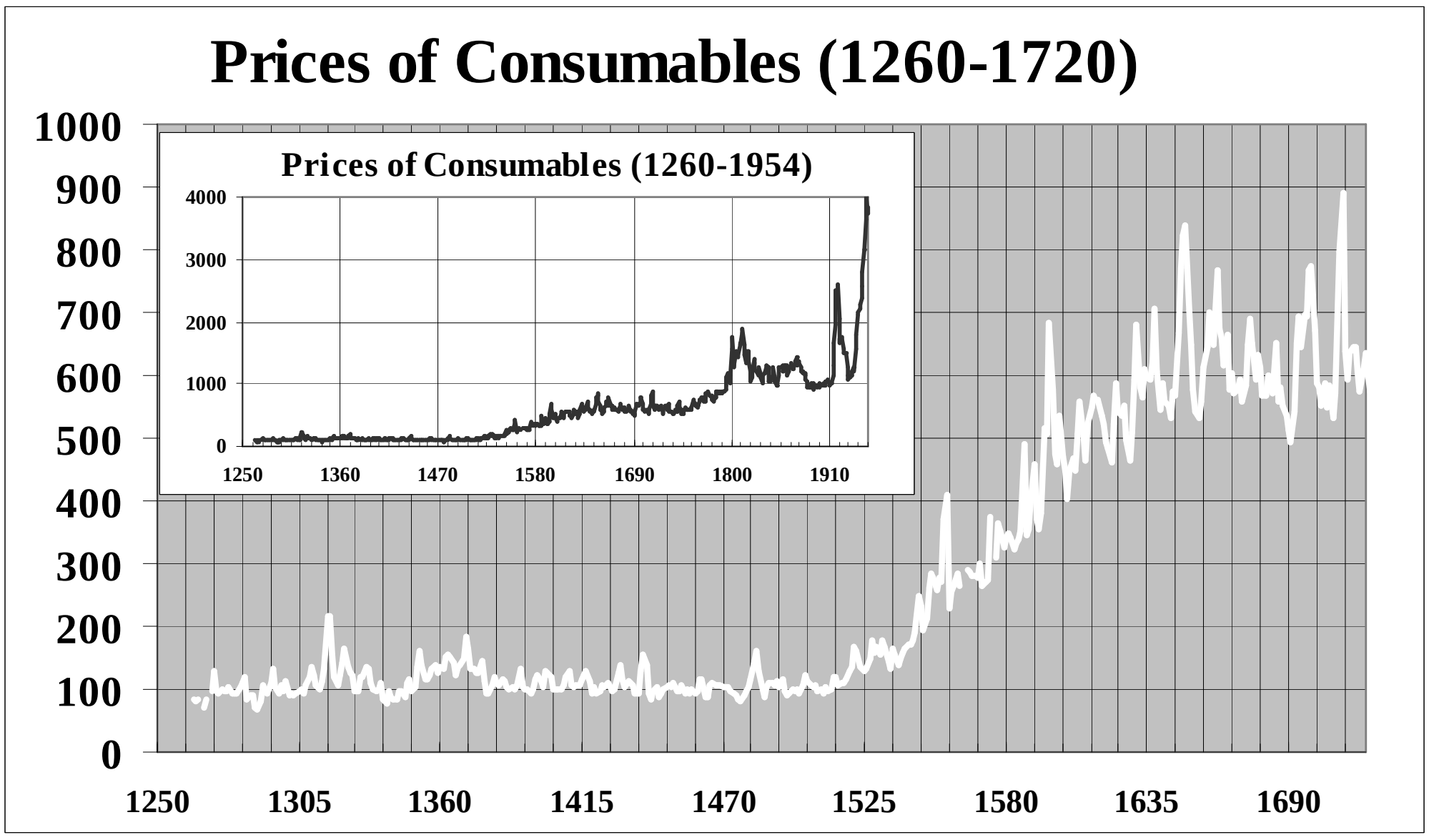

*Fig. 2.* Composite unit of consumables (CUC) index for the studied period (1260 - 1720). These prices are consistent with wheat prices in medieval England reported by Rogers (1887). Inset: the CUC for 1260-1954.

During the next step we repeated the data analysis, as it was made in the first part of our research for wheat prices: restoration of the slow trend component with the following normalization of CUC prices by this slow component gave us relative variations of CUC prices; the noise component was filtered from the burst component by amplitude discrimination (the level of 27.5% was used); the largest CUC price bursts were identified for each 11-year period.

Finally, means, medians and standard deviations of inter-burst time intervals were calculated (Table I). The statistical parameters for three used interval distributions (Composite Unit of Consumables, wheat prices, and "minimum sunspot" states) are very similar, and the hypothesis that all three samples have the same nature (are taken from the same statistical population) cannot be rejected on a 0.01 significance level.

TABLE I.

Comparison of statistical parameters for three studied samples: burst-burst intervals for prices of composite unit of consumables, burst-burst sample for wheat prices, minimum-minimum intervals for sunspot cycle.

| Sample | Median (years) | Average (years) | Standard deviation (years) |
|---|---|---|---|
| Price burst to burst interval according to: | | | |
| Composite Unit of Consumables | 10.0 | 10.65 | 1.57 |
| Wheat prices (1259-1702) | 11.0 | 11.14 | 1.44 |
| Minimum to minimum sunspot intervals (1700-2000) | 10.7 | 11.02 | 1.53 |

Another indication on the common nature of CUC price bursts, wheat prices bursts and sunspot minimum states is illustrated in Fig. 3 where three histograms of the interval distributions for the considered samples are shown. Comparison of the histograms with $\chi^2$-criterion enables accepting of the hypothesis, that they are taken from the same statistical population, at a significance level of 0.05 – 0.10.

In this statistical test we treated the samples of CUC and of wheat prices as independent samples, for the purpose of comparison, because the weight of wheat prices in the CUC index is less than 10%. However, in the reality of medieval England, wheat prices were an essential part of the total costs of consumables, both in a direct manner when wheat was purchased for food, and indirectly, when it was represented in workers' salaries. In some sense, the role of wheat as the main source of muscular energy was similar to the role of petroleum in our times as source of electric energy. In any case, good agreement between interval distributions of consumable price bursts and sunspots confirms our previous conclusion that solar activity influenced wheat prices in medieval England.

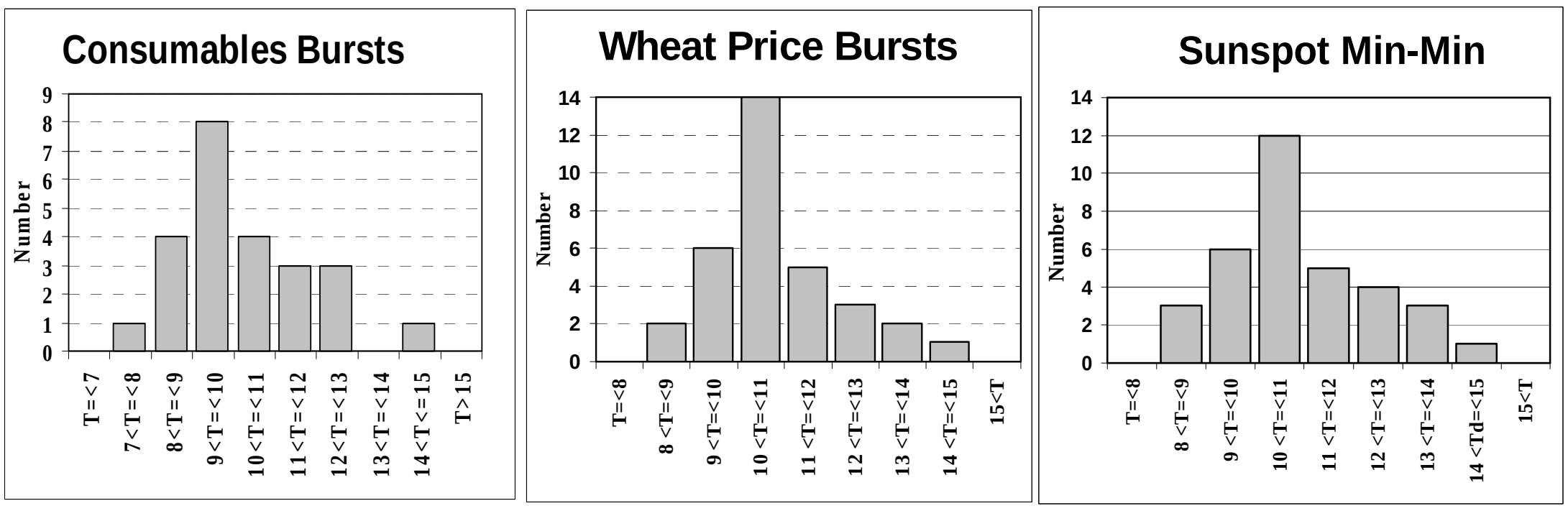

*Fig 3.* Comparison of interval distributions for CUC prices, wheat price bursts, and minimum-to-minimum sunspot intervals.

## 3.2 POSSIBLE MANIFESTATION OF THE SOLAR ACTIVITY IN THE MODERN USA WHEAT MARKET

As discussed in the first part of our work (Pustilnik and Yom Din, 2004), our hypothesis was that solar activity effects in modern times are significantly diminished by three previously negligible effects: (a) technological innovations (e.g., genetic selection, genetic engineering, control of plant diseases) that increase the resistance of cultivated crops to unfavorable environmental conditions such as weather abnormalities; (b) globalization of the world market that protects local markets from unfavorable conditions such as local crop failures; (c) governmental intervention that protects growers against price bursts by paying premiums to decrease/maintain crop area or encouraging purchase of crop insurance. On the other hand, in some situations such compensation mechanisms may be ineffective. While they may be effective in high-risk agriculture in developed countries, in developing regions the lack of financial resources may prevent such countries from benefiting from these advances.

To test the applicability of our approach to modern times, we investigated wheat prices in the USA in 1909-1992 (USDA, 2004) (Fig. 4). Clearly, a small sample that includes only eight sunspot cycles does not enable investigation of the statistical properties of inter-burst intervals. In this situation, we can only test maximum-minimum price asymmetry, such as that discovered for wheat prices in medieval England during the Maunder minimum century in 1600-1700 (Pustilnik and Yom Din, 2004).

To test the Max-Min price asymmetry, we examined wheat price variations in the USA in 1909-1992 (Fig. 4), marked the moments of sunspot maximum and minimum (white triangles and black squares, respectively) and price transition from state of minimal activity to the maximal one (arrows, white for raising of price and black for fall down). In the upper chart in Fig. 4 the relative differences $\Delta\mathrm{Price}$ between prices observed in maximum ( $p_{\max}$ ) and previous minimum ( $p_{\min}$ ) states of solar activity, normalized by the average price $(p_{\max} + p_{\min})/2$, are shown for every min-max cycle. The sample means were estimated as $\overline{\Delta\mathrm{Price}} = 0.29$ and the standard deviation as $s(\overline{\Delta\mathrm{Price}}) = 0.12$. This allowed rejecting the one-tailed zero hypothesis about the non-positive mean value of the price difference on a significance level $\alpha < 0.05$. Thus, it can be accepted that the Max-Min price asymmetry for the studied sample does exists. We wish to note that the amplitude and significance of this asymmetry are lower than those measured for wheat prices in medieval England in the period of the Maunder minimum. This is not surprising considering the globalization of the USA – UK wheat market documented beginning from the second half of the 19$^{th}$ century (Fremdling 1999).

The studied wheat market was influenced by major political and economic cataclysms: two World Wars (1914-1921, 1939-1945) and the Great Depression (1929-1941). The existence of a significant maximum-minimum price asymmetry in spite of these disruptions and suppression effects described above could not be expected a priori. A possible explanation of this surprising result is the compact localization of wheat production in USA (especially, durum and spring wheat). For example, about 70% of all durum of the USA is produced in a part of North Dakota whose area is less than 2% of the USA. Clearly, a high concentration of the crop area in so small a region increases sensitivity of wheat production to weather abnormalities and, among them, abnormalities caused by space weather.

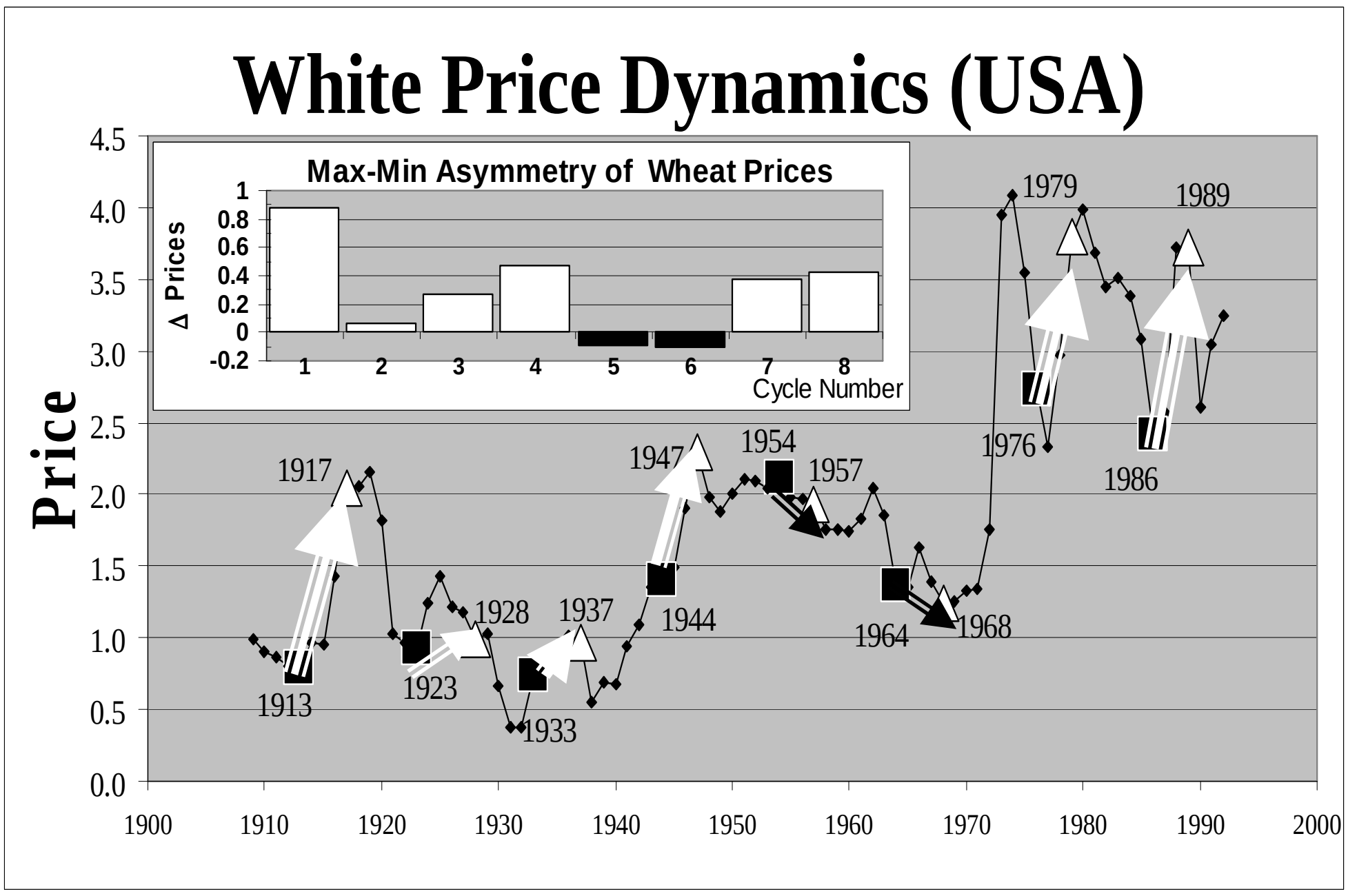

*Fig. 4.* Maximum-minimum price asymmetry for USA wheat prices 1909-1992 presented in dollars per bushel. White triangles are prices during periods of maximum sunspots; black squares are prices during periods of minimum sunspots. White arrows indicate rise in price from minimum to maximum sunspot periods; black arrows indicate drop in price. Inset shows ΔPrice - the relative price differences for each of the eight min-max sunspot cycles: 1913-17, 1923-28, 1933-37, 1944-47, 1954-57, 1964-68, 1976-79, 1986-89. The source of the moments of maximums and minimums is NOAO database (ftp://ftp.ngdc.noaa.gov/STP/SOLAR_DATA/SUNSPOT_NUMBERS/YEARLY).

## 4. Conclusions

1. The test of the interval distribution of the prices of consumables for Medieval England shows a good consistence with the interval distribution of sunspot minimum-minimum. It confirms our previously reported conclusions about the manifestation of the influence of solar activity on wheat prices in that period in the same region.
2. The test of the maximum-minimum price asymmetry for wheat in the USA in the 20-th century shows that the effect of the influence of solar activity also occurred, but its amplitude and its significance level were lower than that for Medieval England in the century of the Maunder Minimum.